\documentclass [12pt] {article}
\usepackage{graphicx}
\newcommand{\cs}[3]{{{#3} \brace {#1 #2}}}
\newcommand{\tder}[2]{\frac{d{#1}}{d{#2}}}
\newcommand{\pder}[2]{\frac{\partial{#1}}{\partial{#2}}}

\newcommand{\h}[1]{\mathop{\lambda}\limits_{#1}\ \!\!\!}

\newcommand{\edf}{\ {\mathop{=}\limits^{\rm def}}\ }

\newcommand{\al}{\alpha}
\newcommand{\be}{\beta}
\newcommand{\m}{\mu}
\newcommand{\n}{\nu}
\newcommand{\s}{\sigma}
\newcommand{\g}{\gamma}

\begin{document}
\title{\bf Quantum Roots in Geometry: I}

\author{{\bf M. I. Wanas}\\
\normalsize Astronomy Department, Faculty of Science, Cairo
university {\bf Egypt}\\ e-mail wanas@frcu.eun.eg} \maketitle
\begin{abstract}
In the present work, it is shown that the geometerization
philosophy has not been exhausted. Some quantum roots are already
built in non-symmetric geometries. Path equations in such
geometries give rise to spin-gravity interaction. Some
experimental evidences (the results of the COW-experiment)
indicate the existence of this interaction. It is shown that the
new quantum path equations could account for the results of the
COW-experiment. Large scale applications, of the new path
equations, admitted by such geometries, give rise to tests for the
existence of this interaction on the astrophysical and
cosmological scales. As a byproduct, it is shown that the quantum
roots appeared explicitly, in the path equations, can be diffused
in the whole geometry using a parameterization scheme.
\end{abstract}
\section{Introduction}
Most of the success of physics in the 20th century has been
achieved as a result of the applications of two philosophies.The
first is the {\it Quantization Philosophy} and the second is the
{\it Geometerization Philosophy}. The consequences of applying the
first is the {\it Quantum Theory}, while the consequences of
applying the second is the {\it General Theory of Relativity},
(GR). The study and understanding of the four known fundamental
interactions are not equally successful using, only, one of these
two rival philosophies. Electromagnetism, weak and strong
interactions are well understood using the quantization
philosophy, while gravity is not understood using this philosophy.
In the context of geometerization of physics, GR is considered as
a good theory for gravity, while there are no such successful
geometric theories for the other three interactions.

It seems that a third philosophy is needed to unify the physics of
the four fundamental interactions. This philosophy may lead to new
physics. This would be, undoubtedly, a difficult task. It would be
of importance to reach the conclusion that the two rival
philosophies are completely exhausted, before trying a third one.
This my be a less difficult task. It needs a careful examination
of applying the existing philosophies. Examination of the
geometric approach to physics shows that this approach is {\bf
not} exhausted yet. Some types of geometry admit some quantum
properties. This is what I am going to show in the present work.

The following statement summarizes the philosophy of
gemeterization of physics:
\begin{center}
{\it "To understand nature one should start with geometry and end
with physics"}.
\end{center}
In applying this philosophy, one should look for an appropriate
geometry. Einstein, in applying his geometerization philosophy,
used three types of geometry. Some of the main properties of these
geometries are summarized in the following table.
\begin{center}
 Table I: Comparison between 3-types of geometry\\
 %\vspace{0.5cm}
 \begin{tabular}{|c|c|c|c|} \hline
 & & &  \\ Geometry [Ref.] & Metric & Connection &  Building
 Blocks (\#)
 \\ & & &
 \\ \hline & & &  \\ Riemannian [1] & Symmetric & Symmetric &
 Metric tensor (10)
 \\ & & &  \\ \hline & & &  \\ Absolute Parallelism [2] & Symmetric &
 Non-symmetric &
 Tetrad vectors (16)
 \\ & & &  \\ \hline & & &  \\ Einstein Non-symmetric [3] & Non-symmetric &
 Non-symmetric
 & Metric tensor (16)\\
 & & &  \\ \hline
 \end{tabular}
 \end{center}
We mean by the term {\it "Building Block"} the geometric object,
using which one can construct the whole geometry. In the last
column of Table I, we assume that the dimension of space $ n = 4.$

Riemannian geometry has been used by Einstein to construct his
successful theory of gravity, GR. It is well known that the number
of building blocks in this geometry is just sufficient to describe
gravity. For this reason, we are going to consider the other two
geometries, in Table I, since the number of building blocks in
each is enough to accommodate other interactions, together with
gravity. These interactions may have some quantum properties.

The term {\it "Non-Symmetric Geometry"} will be used to indicate
that the geometry admits  non-symmetric connection. In such a
geometry, one can define define three types of tensor derivatives
(derivatives that preserve tensor properties):
$$
A^{\mu}_{+|~ \nu} {\stackrel{def.}{=}}~ A^{\mu}_{~, \nu} +
A^{\alpha}C^\mu _{.\alpha \nu} ,\eqno{(1.1)}
$$
$$
A^{\mu}_{-|~ \nu} {\stackrel{def.}{=}}~ A^{\mu}_{~, \nu} +
A^{\alpha}C^\mu_{.\nu \alpha} ,\eqno{(1.2)}
$$
$$
A^{\mu}_{0|~ \nu} {\stackrel{def.}{=}}~ A^{\mu}_{~, \nu} +
A^{\alpha}C^\mu _{.(\alpha \nu)} ,\eqno{(1.3)}$$
 where $A^\mu$ is
an arbitrary vector and $C^\mu_{.\nu\al}$ is the
 non-symmetric connection. Braces ( ) are used for symmetrization and
brackets [ ] will be used for anti-symmetrization. The comma is
used for ordinary (not tensor) partial differentiation.

Now, what is the starting point for examining non-symmetric
geometries to look for any quantum features? It is well known
that, quantum properties in microscopic world were discovered when
Planck tried to interpret black body radiation, a phenomena which
is closely connected to the motion of electrons. On the other
hand, in the context of geometerization of physics, motion is
described using paths (curves) of an appropriate geometry. So, a
good starting point, may be a search for path equations in the
geometries under consideration.

Bazanski [4],[5] has established a new approach to derive the
equations of geodesic and geodesic deviation simultaneously by
carrying out  variation on the following Lagrangian:
$$
L_{B} = g_{\mu\nu} U^{\mu} {\frac{D {\Psi}^{\nu}}{D S}}, \eqno{(1.4)}
$$
where $U^{\mu} \edf {\tder{x^{\mu}}{S}}$,  $g_{\mu\nu}$ is the
metric tensor, $\Psi^{\mu}$ is the deviation vector and
$\frac{D}{D S}$ is the covariant differential operator using
Christoffel Symbol. We are going to generalize the Bazanski
approach, by replacing the covariant derivative, used in his
Lagrangian, by tensor derivatives of the types given by (1.1),
(1.2) and (1.3), admitted by the geometry under consideration.

The work in this review is organized as follows: Section 2 gives a
brief review of  the two non-symmetric geometries under
consideration, together with the new path equations resulting from
each one. Section 3 gives some remarks about the quantum features
appearing in these geometric paths. A method for diffusing the
quantum properties, in the whole geometry, is given in Section 4.
The general quantum path, of the absolute parallelism geometry, is
linearized in Section 5. Section 6 gives confirmation and
applications of the quantum paths. the work is discussed and
concluded in Section 7.

\section{ Geometries with Built-in Quantum Roots }

\subsection{The Absolute Parallelism Geometry}

  Absolute parallelism (AP)space is an n-dimensional manifold each
 point of which is labelled by n-independent variables $x^{\nu} (\nu =1,2,3,...,n)$
 and at each point we define n-linearly independent contravariant vectors
$\h{i}^\mu (i=1,2,3,3...,n$, denotes the vector number and $\mu
=1,2,3...,n$, denotes the coordinate component) subject to the
condition, $$
 \h{i}^{\stackrel{\m}{+}}_{.|~ \nu}=0,  \eqno{(2.1)}
$$
 where the stroke and the (+) sign denote absolute differentiation, using
 a non-symmetric connection  to be defined later. Equation (2.1)is the condition
 for the absolute parallelism.
 The covariant components of $\h{i}^\mu$ are defined such that
 $$
\h{i}^\mu \h{i}_\nu = \delta^{\mu}_{\nu}, \eqno{(2.2)}
 $$
and
 $$
  \h{i}^{\nu} \h{j}_{\nu} = \delta_{ij}.  \eqno{(2.3)}.
$$ Using these vectors, the following second order symmetric
tensors are defined: $$ g^{\mu \nu} \edf \h{i}^\mu \h{i}^{\nu},
\eqno{(2.4)} $$

$$ g_{\mu \nu} \edf \h{i}_\mu \h{i}_{\nu}, \eqno{(2.5)} $$
consequently,
 $$ g^{\mu \alpha} g_{\nu \alpha} = \delta^{\mu}_{\ \nu }. \eqno{(2.6)} $$
These second order tensors can serve as the metric tensor and its
conjugate of Riemannian space, associated with the AP-space, when
needed. This type of geometry admits, at least, four affine
connections. The first is a non-symmetric connection given as a
direct solution of the AP-condition(2.1), i.e. $$
\Gamma^{\alpha}_{.~\mu \nu} = \h{i}^{\alpha} \h{i}_{\mu,\nu}.
\eqno{(2.7)} $$ The second is its dual $
\hat{\Gamma}^{\alpha}_{.~\mu \nu}( = \Gamma^{\alpha}_{.~\nu \mu}) $,
since (2.7) is non-symmetric. The third one is the symmetric part of
(2.7),  $ \Gamma^{\alpha}_{.(\mu \nu)}$. The fourth is Christoffel
symbol defined using (2.4),(2.5) ( as a result of imposing a
metricity condition). The torsion tensor is the skew symmetric part
of the affine connection (2.7), i.e.  $$ \Lambda^{\alpha}_{.~\mu
\nu} \edf \Gamma^{\alpha}_{.~\mu \nu} - \Gamma^{\alpha}_{.~\nu \mu}.
\eqno{(2.8)} $$ Another third order tensor (contortion) is defined
by, $$ \gamma^{\alpha}_{.~\mu \nu} \edf \h{i}^{\alpha} \h{i}_{\mu ;
\nu}. \eqno{(2.9)} $$ The semicolon is used to characterize
covariant differentiation using Christoffel symbol. The two tensors
are related by the formula,
$$\gamma^{\alpha}_{.\mu \nu}= \frac{1}{2} (\Lambda^{\alpha}_{.\mu
\nu } - \Lambda^{~ \alpha}_{\nu .\mu} - \Lambda^{~\alpha}_{\mu
.\nu}). \eqno{(2.10)}
 $$ A basic vector could be
 obtained by contraction of one of the above third order tensors,
 i.e.
$$ C_{\mu} \edf  \Lambda^{\alpha}_{.\mu \alpha }=
\gamma^{\alpha}_{. \mu \alpha}. \eqno{(2.11)}$$

 The curvature tensor of the AP-space is, conventionally,  defined by,
 $$
B^{\al}_{.\m \n \s} {\ }  \edf {\ } \Gamma^{\al}_{.\m  \s, \n} -
\Gamma^{\al}_{.\m \n, \s} + \Gamma^{\al}_{\epsilon \n} \Gamma^
{\epsilon}_{. \m \s} - \Gamma^{\al}_{. \epsilon \s}
\Gamma^{\epsilon}_{. \m \n} \equiv 0.\eqno{(2.12)}
 $$
This tensor vanishes identically because of (2.1).

The autoparallel paths, of this geometry, are given by the
equation,
$$ \frac{d^{2}x^{\mu}}{d \lambda^{2}}+ \Gamma^{\mu}_{\alpha \beta}
\frac{d x^{\alpha} }{d \lambda}\frac{d x^{\beta} }{d \lambda} =
0.\eqno{(2.13)} $$

 The AP-geometry, in its conventional form, has two main problems concerning
applications: The first is the identical vanishing of its
curvature tensor and the second is that its path equations (2.13)
do not represent any known physical trajectory. These problems
will be treated in Section 4.

Many authors believe that, because of (2.12), the AP-space is flat
. It is shown [6] that AP-spaces are, in general, curved. The
problem of curvature in AP-spaces is a problem of definition. In
any affinely connected space there is, at least, two methods for
defining the curvature tensor. The first method is by replacing
Christoffel symbol, in the definition of Riemann-Christoffel
tensor, by the affine connection defined in the space concerned.
The second method is to define curvature as a measure of
non-commutation of tensor differentiation using the affine
connection of the space. It is known that, the two methods give
identical results in case of Riemannian space. But the situation
is different for spaces with non-symmetric connections. The two
methods are not identical.

 The application of the second method
in non-symmetric geometries implies a problem. That is, we usually
use an arbitrary vector in order to study the non-commutation of
tensor differentiation, and the resulting expression will not be
free from this vector. Fortunately, this problem is solved in
AP-spaces [7]. We can replace the arbitrary vector by the vectors
defining the structure of AP-spaces. In this case we can define the
following curvature tensors (I am going to call these tensors
\textit{non-conventional curvature tensors}):

$$ \h{i}^{\stackrel{\m}{+}} _{\  | {\n \s}}  -
\h{i}^{\stackrel{\m}{+}}_ {\  | {\s \n}}   = \h{i}^\al B^\m_{. \al
\n \s},\eqno{(2.14)} $$ $$ \h{i}^{\stackrel{\m}{-}}_ {\ | {\n \s}}
- \h{i}^{\stackrel{\m}{-}}_{\  | {\s \n}}   = \h{i}^\al L^\m_{.
\al \n \s},\eqno{(2.15)} $$ $$ \h{i}^{\stackrel{\m}{0}}_ {\ | {\n
\s}}  - \h{i}^{\stackrel{\m}{0}} _{\ | {\s \n}}   = \h{i}^\al
N^\m_{. \al \n \s},\eqno{(2.16)} $$
 $$
\h{i}^{\mu}_{\  ; \n \s}  - \h{i}^{\mu}_{\  ; \s \n}   = \h{i}^\al
R^\m_{. \al \n \s} ,\eqno{(2.17)}  $$ here we use the stroke , a (+)
sign  and (-) sign to characterize absolute differentiation using
the connection (2.7) and its dual, respectively. We use the stroke
without signs to characterize absolute differentiation using the
symmetric part of (2.7), while the semicolon is used to characterize
covariant differentiation using the Christoffel symbols. The
non-conventional curvature tensors defined by (2.14), (2.15), (2.16)
and (2.17) are in general non-vanishing except the first one, which
vanishes (because of the AP-condition (2.1)).

The non-conventional curvature tensors defined above can be written
explicitly in terms of torsion, or contortion via (2.10), i.e.  $$
B^{\alpha}_{. \mu \nu \sigma} = R^{\alpha}_{. \mu \nu \sigma} +
Q^{\alpha}_{. \mu \nu \sigma} \equiv 0 ,\eqno{(2.18)} $$ $$
 L^{\alpha}_{. \mu \nu \sigma} \edf \Lambda^{\stackrel{\alpha}{+}}_{. {\stackrel{\m}{+}} {\stackrel{\nu}{-}}
 | \sigma} - \Lambda^{\stackrel{\alpha}{+}}_{. {\stackrel{\m}{+}} {\stackrel{\sigma}{-}}
 | \nu} + \Lambda^{\beta}_{. \mu \nu} \Lambda^{\alpha}_{. \sigma
 \beta} - \Lambda^{\beta}_{. \mu \sigma} \Lambda^{\alpha}_{. \nu
 \beta},
 \eqno{(2.19)}$$
 $$
N^{\alpha}_{. \mu \nu \sigma} \edf \Lambda^{\alpha}_{. \mu \nu |
\sigma } - \Lambda^{\alpha}_{. \mu \sigma | \nu } +
\Lambda^{\beta}_{. \mu \nu}\Lambda^{\alpha}_{.  \beta \sigma} -
\Lambda^{\beta}_{. \mu \sigma}\Lambda^{\alpha}_{.  \nu \beta},
\eqno{(2.20)}
 $$
$$ Q^{\alpha}_{. \mu \nu \sigma} \edf
\gamma^{\stackrel{\alpha}{+}}_{. {\stackrel{\m}{+}}
{\stackrel{\nu}{+}}
 | \sigma} - \gamma^{\stackrel{\alpha}{+}}_{. {\stackrel{\m}{+}} {\stackrel{\sigma}{-}}
 | \nu} + \gamma^{\beta}_{. \mu \sigma} \gamma^{\alpha}_{. \beta
 \nu} - \gamma^{\beta}_{. \mu \nu} \gamma^{\alpha}_{.  \beta \sigma},
 \eqno{(2.21)}
$$ It is clear that the vanishing of the torsion will lead to the
vanishing of (2.19), (2.20). Also this will lead to vanishing of
(2.21) via (2.10) and consequently the vanishing of $
R^{\alpha}_{. \mu \nu \sigma}$ via (2.18). This represents another
problem facing field theories written in AP-spaces. Such theories
will not have GR limit as the torsion vanishes, if this condition
is needed.

\subsection{Quantum Properties of the AP-Geometry}
Recently [8], using the affine connections defined in the AP-space
to generalize the Bazanski Lagrangian (1.4),  three path equations
were discovered in the AP-geometry . These equations can be
written in the form:
$$ {\frac{dU^\m}{dS^-}} + \{^{\m}_{\al\be}\} U^\al U^\be = 0,
\eqno{(2.22)}$$
 $$ {\frac{dW^\m}{dS^0}} + \{^{\m}_{\al\be}\} W^\al W^\be = -
{\frac{1}{2}} \Lambda^{~ ~ ~ ~ \m}_{(\al \be) .}~~ W^\al W^\be,
\eqno{(2.23)} $$ $$ {\frac{dV^\m}{dS^+}} + \{^{\m}_{\al\be}\} V^\al
V^\be = - \Lambda^{~ ~ ~ ~ \m}_{(\al \be) .} ~~V^\al V^\be,
\eqno{(2.24)} $$ where $ S^{-}$,$ S^{0}$and $ S^{+}$ are the
parameters varying along the corresponding curves whose tangents
are ${ J}^\alpha$,$ W^\alpha$ and ${ V}^\alpha$, respectively. We
can write the new set of the path equations, obtained in this
geometry, in the following form:
$$
{\frac{dB^\mu}{d\hat S}} + a~ \cs{\alpha}{\beta}{\mu}\ B^{\alpha}
B^{\beta} = - b~ \Lambda_{(\alpha \beta)}^{~.~.~~\mu}~~B^{\alpha}
B^{\beta}, \eqno{(2.25)}
$$
where $a$, $b$ are the numerical coefficients of the Christoffel
symbol term and of the torsion term, respectively. Thus we can
construct the following table.

\begin{center}
 Table II: Numerical Coefficients of The Path Equation in AP-Geometry      \\
%\vspace{0.5cm}
\begin{tabular}{|c|c|c|} \hline
& & \\
Affine Connection Used &Coefficient $a$  &Coefficient $b$   \\
& & \\ \hline
& & \\
${\hat\Gamma}^{\alpha}_{.~\mu \nu}$     & 1   & 0  \\
& & \\ \hline
& & \\
${\Gamma}^{\alpha}_{.~(\mu\nu)}$      &  1   & $\frac{1}{2}$ \\
& & \\ \hline
& & \\
${\Gamma}^{\alpha}_{.~\mu \nu}$ &1 &1 \\
& & \\ \hline

\end{tabular}
\end{center}
The first column in this table contains the affine connections
used to generalize the Bazanski Lagrangian.
The set of equations (2.22), (2.23)and (2.24) possesses some interesting features: \\
1. It gives the effect of the torsion on the curves (paths)of the
geometry. \\ 2. This set is irreducible i.e. no one of these
equations can be reduced to the other unless the torsion vanishes.
This vanishing will lead to flat space (in view of the definitions
(2.18-21)), which is not
 suitable for applications.\\
3. The coefficient of the torsion term jumps by a step of one-half
from one equation to the next (as clear from
Table II).\\
The last feature is tempting to conclude that:

$\underline{ "paths\ in\ AP-geometry\ are\ naturally\
quantized"}$.
\subsection{Einstein Non-symmetric Geometry}
Einstein generalized the Riemannian geometry by dropping the
symmetry conditions imposed on the metric tensor and on the affine
connection [3]. In this geometry the non-symmetric metric tensor
is given by:
$$g_{\mu \nu} {\stackrel{def.}{=}}~ h_{\mu \nu} + f_{\mu \nu} ,\eqno{(2.26)}
$$
where, $$
        h_{\mu \nu} {\stackrel{def.}{=}}~ \frac{1}{2}( g_{\mu \nu} + g_{\mu \nu})   ,
$$
$$
     f_{\mu \nu} {\stackrel{def.}{=}}~ \frac{1}{2}(g_{\mu \nu}  -  g_{\mu \nu})   .
$$
Since the connection of the geometry, $U^{\alpha}_{.~\mu \nu}$, is
assumed to be non-symmetric, one can define the following 3-types of
covariant derivatives:
$$
A^{\mu}_{+||~ \nu} {\stackrel{def.}{=}}~ A^{\mu}_{~, \nu} +
A^{\alpha}U^\mu _{.\alpha \nu} ,\eqno{(2.27)}
$$
$$
A^{\mu}_{-||~ \nu} {\stackrel{def.}{=}}~ A^{\mu}_{~, \nu} +
A^{\alpha}U^\mu_{.\nu \alpha} ,\eqno{(2.28)}
$$
$$
A^{\mu}_{0||~ \nu} {\stackrel{def.}{=}}~ A^{\mu}_{~, \nu} +
A^{\alpha}U^\mu _{.(\alpha \nu)} ,\eqno{(2.29)}
$$
where $A^\mu$ is any arbitrary vector. Now the connection
$U^{\alpha}_{.~\mu \nu}$ is defined such that [3],
$$
g_{\stackrel{\mu}{+} \stackrel{\nu}{-} ||\sigma} = 0 ,\eqno{(2.30)}
$$
$$i.e.~~~ g_{\mu \nu ,\sigma} = g_{\mu \alpha} U^{\alpha}_{.~~\sigma
\nu} + g_{ \alpha \nu} U^{\alpha}_{.~~\mu \sigma}. \eqno{(2.31)}
$$
The non-symmetric connection can be written in the the following
form:
$$U^{\alpha}_{.~~\mu \nu} {\stackrel{def.}{=}}~ U^{\alpha}_{.~(\mu
\nu )} + U^{\alpha}_{.~~[\mu \nu]} =  \cs{\mu}{\nu}{\alpha}\ +
K^{\alpha}_{.~~\mu \nu}, \eqno{(2.32)}
$$
where,
$$U^{\alpha}_{.~(\mu \nu)} {\stackrel{def.}{=}}~
\frac{1}{2}(~U^{\alpha}_{.~~\mu \nu } + U^{\alpha}_{.~~\nu \mu })
, \eqno{(2.33)}
$$
$$U^{\alpha}_{.~~[\mu \nu]} {\stackrel{def.}{=}}~
\frac{1}{2}(~U^{\alpha}_{.~\mu \nu } - U^{\alpha}_{.~ \nu \mu}) =
K^{\alpha}_{.~[\mu \nu]} = \frac{1}{2} S^{\alpha}_{.~\mu \nu}   ,\eqno{(2.34)}
$$
where $S^{\alpha}_{.~\mu \nu}$ is a third order tensor
representing the torsion of the Einstein non-symmetric (ENS)
geometry.

  The contravariant metric tensor $g^{\mu \nu}$ is defied such that :
$$
g^{\mu \alpha}g_{\nu \alpha} = g^{\alpha \mu}g_{\alpha \nu} =
\delta^{\mu}_{\nu}  .\eqno{(2.35)}
$$
The tensor derivatives {(2.27), (2.28) and (2.29)} are connected to
the parameter derivatives by the relations :
$$
\frac{\nabla A^\mu}{\nabla \tau^{-}} = A^{\mu}_{-||~
\alpha}\tilde{J}^\alpha, \eqno{(2.36)}
$$
$$\frac{\nabla A^\mu}{\nabla \tau^{0}} = A^{\mu}_{0||~
\alpha}\tilde{W}^\alpha, \eqno{(2.37)}
$$
$$
\frac{\nabla A^\mu}{\nabla \tau^{+}} = A^{\mu}_{+||~
\alpha}\tilde{V}^\alpha ,\eqno{(2.38)}
$$
where $\tilde{J}^{\mu}$,$\tilde{W}^{\mu}$ and $\tilde{V}^{\mu}$
are tangents to the paths whose evolution parameters are
$\tau^{-}$,$\tau^{0}$ and $\tau^{+}$, respectively.\\

\subsection{Quantum Properties of ENS-Geometry}
  Applying the Bazanski approach
 to the Lagrangian functions:
$$
\Xi^{-} = g_{\mu \alpha}\tilde{J}^{\mu}\frac{\nabla
\Psi^\alpha}{\nabla\tau^{-}} ,\eqno{(2.39)}
$$
$$
\Xi^{0} = g_{\mu \alpha}\tilde{W}^{\mu}\frac{\nabla
\Theta^\alpha}{\nabla\tau^{0}} ,\eqno{(2.40)}
$$
$$\Xi^{+} = g_{\mu \alpha}\tilde{V}^{\mu}\frac{\nabla
\Phi^\alpha}{\nabla\tau^{+}} ,\eqno{(2.41)}
$$
where $\Psi^\alpha, \Theta^\alpha$ and $\Phi^\alpha$ are the
deviation vectors, we get [9] the following set path equations
respectively,
$${\frac{d\tilde{J}^\alpha}{d\tau^{-}}} + \cs{\mu}{\nu}{\alpha}\
\tilde{J}^{\mu}\tilde{J}^{\nu} =
-K^{\alpha}_{.~\mu\nu}\tilde{J}^{\mu}\tilde{J}^{\nu} ,\eqno{(2.42)}
$$
$$
{\frac{d\tilde{W}^{\alpha}}{d\tau^{0}}} + \cs{\mu}{\nu}{\alpha}\
\tilde{W}^{\mu}\tilde{W}^{\nu} =  -\frac{1}{2} g^{\alpha
\sigma}g_{\mu \rho}S^{\rho}_{.~\nu \sigma}
\tilde{W}^{\mu}\tilde{W}^{\nu} -
K^{\alpha}_{.~\mu\nu}\tilde{W}^{\mu}\tilde{W}^{\nu} ,\eqno{(2.43)}
$$
$$
{\frac{d\tilde{V}^\alpha}{d\tau^{+}}} + \cs{\mu}{\nu}{\alpha}\
\tilde{V}^{\mu}\tilde{V}^{\nu} =
 - g^{\alpha \sigma}g_{\mu \rho}S^{\rho}_{.~\nu \sigma} \tilde{V}^{\mu}\tilde{V}^{\nu} - K^{\alpha}_{.~\mu\nu}\tilde{V}^{\mu}\tilde{V}^{\nu}.\eqno{(2.44)}
$$
This set of equations can be written in the following general
form:
$${
\frac{dC^\alpha}{d\tau} +a~ \cs{\mu}{\nu}{\alpha}\ C^{\mu}C^{\nu}
= - b~ g^{\alpha \sigma} g_{\mu \rho} S^{\rho}_{.~\nu \sigma}
C^{\mu}C^{\nu} - c~ K^{\alpha}_{.~\mu\nu} C^{\mu} C^{\nu}}.\eqno{(2.45)}
$$
where $a$, $b$ and $c$ are the numerical coefficient of the
Christoffel symbol, torsion and K-terms, respectively. Thus, we can
construct the following table:
\begin{center}

  Table III: Coefficients of The Path Equations in ENS-Geometry \\
%\vspace{0.5cm}
\begin{tabular}{|c|c|c|c|} \hline
Affine Connection used &Coefficient $a$        &Coefficient $b$     &Coefficient $c$   \\
\hline
& & & \\
${\hat U}^{\alpha}_{.~\mu \nu}$    & 1   & 0  & 1 \\
& & & \\ \hline
& & & \\
$U^{\alpha}_{.~(\mu\nu)}$       & 1   & $\frac{1}{2}$  &1  \\
& & & \\ \hline
& & & \\
$U^{\alpha}_{.~\mu \nu}$
&1    & 1      &1  \\
& & & \\ \hline
 \end{tabular}
\end{center}
The first column in this table contains the affine connections
used to generalize the Bazanski Lagrangian.

From this table, it is clear that, the jumping coefficient of the
torsion term (column 3) has the same values obtained in the case
of the AP-geometry (Table II, column 3). So, one can draw a similar
conclusion given in Subsection 2.2:

$\underline{ "Paths\ in\ ENS-geometry\ are\ naturally\
quantized"}$

\section{Features of Quantum Roots}

(i) We consider the jump of the coefficient of the torsion term in
the path equations of Subsections 2.2 and 2.4, by a step of
one-half, as quantum roots emerging in non-symmetric geometries.
Such path equations,  are usually used to represent trajectories of
test particles, in the context of the scheme of geometerization of
physics. So, if such trajectories do exist in nature, then one can
conclude that space-time is quantized and the geometry describing
nature should be non-symmetric.

(ii) The quantum properties shown in Tables II and III, are
properties built in the examined geometries. In other words, these
properties are intrinsic properties characterizing the type of
geometry used. The properties mentioned above are not consequences
of applying any known quantization schemes.

(iii) In the scheme performed to discover these properties, certain
Lagrangian functions are used. Such functions contain, in their
structure, covariant derivatives, in which certain affine
connections are used. The quantum properties discovered are
closely related to such connections. It is well known that, in any
non-symmetric geometry, one can define more affine connections by
adding any third order tensor to any affine connection already
defined in the geometry. If we do so, in the geometries examined
in Section 2, one would not get any values (for the coefficients
given in Tables II, III) different from those listed in the two
tables. As a check one can try the connection,

 $$ \Omega^{\alpha}_{. \mu\nu}
 \edf \{^{\alpha}_{\mu\nu}\} + \Lambda^{\alpha}_{. \mu \nu},
\eqno{(3.1)} $$ defined in the AP-geometry.

(iv) As stated above, the quantum properties discovered are
closely connected to the affine connection, or more strictly, to
its skew pare, the torsion tensor. The coefficients of Christoffel
symbol term (the second column of Tables II and III) are the same
for all paths. Also, the coefficient of the symmetric part of the
tensor $K^a _{\m \n}$ has no such jumping properties (last column of
Table III).

\section{Parameterization and Diffusion of Quantum Roots}
It is now obvious that the quantum roots discovered in non-symmetric
geometries depend mainly on the existence of non-symmetric
connections admitted by such geometries. Furthermore, these roots,
explicitly, appeared first in the path equations and not in  other
geometric entity.

In order to extend these roots to the whole geometry, we are going
to reconstruct the geometry using a general affine connection. This
connection is defined as a linear combination of the connections,
already, admitted by the geometry. The combination is carried out
using certain parameters. The general expression obtained may not
represent an affine connection, in a conventional sense. In other
words, it might not be transformed as an affine connection, under
the group of general coordinate transformation, unless certain
conditions are imposed on the values of the
 parameters used. The version of the geometry obtained
in this way is a {\it parameterized} version.

In the case of the AP-geometry, using the affine connections mentioned in
Subsection 2.1 and carrying out the parameterization scheme mentioned above,
the following results are obtained [10]:
 Combining linearly the above mentioned connections we get, after some reductions,
the following parameterized expression, $$ \nabla^{\mu}_{. \alpha
\beta } = (a+b) \{^{\mu}_{\alpha\beta}\} + b  \gamma^{\mu}_{.
\alpha \beta } \eqno{(4.1a)} $$ where $a$ and $b$ are parameters.
As a consequence of imposing a metricity condition, using (4.1a),
we get
 $$ a + b = 1. \eqno{(4.1b)}$$So, expression (4.1a) will reduce to,
$$ \nabla^{\mu}_{. \alpha
\beta } =  \{^{\mu}_{\alpha\beta}\} + b  \gamma^{\mu}_{. \alpha
\beta },\eqno{(4.2)} $$ which is a general parameterized affine
connection. Using (4.2) to generalize the Bazanski Lagrangian
(1.4), we get
$$ {\frac{dZ^\m}{d\tau}} + \{^{\m}_{\al\be}\} Z^\al
Z^\be = - b~~ \Lambda^{~ ~ ~ ~ \m}_{(\al \be) .} ~~Z^\al Z^\be,
\eqno{(4.3)} $$ where
 ${\tau}$ is a parameter varying along the path and ${Z^{\mu}}$ is the tangent to the path.

 All curvature tensors defined in this parameterized version of
 geometry, are non-vanishing. For example if we redefine the curvature (2.12) using the
 connection (4.2) we get [10] $$  {B^{\ast}}^{\alpha}_{.\mu \nu \sigma} =  R^{\alpha}_{. \mu \nu \sigma} + b~~ \hat Q^{\alpha}_{. \mu \nu \sigma }.  \eqno{(4.4)}$$
where $R^{\alpha}_{.\mu \nu \sigma}$ is Riemann-Christoffel
curvature tensor and $Q^{\alpha}_{.\mu \nu \sigma}$ is defined by,
$$
 \hat Q^\alpha_{.\ \mu \nu \sigma} \edf {\gamma}^{\stackrel{\al}{+}}_{{.\
{\stackrel{\mu}{+}}{\stackrel{\nu}{+}}} | \sigma} -
{\gamma}^{\stackrel{\alpha}{+}}_{.\
{{\stackrel{\mu}{+}}{\stackrel{\sigma}{-}}} | \nu} + b~(
{\gamma}^{\beta}_{. \mu \sigma} {\ } {\gamma}^{\alpha}_{. \beta
\nu} {\ } - {\gamma}^{\beta}_{. \mu \nu} {\ } {\gamma}^{\alpha}_{.
\beta \sigma}). \eqno{(4.5)}
$$

This tensor is, in
 general non-vanishing although the corresponding one (2.18) vanishes identically in the
 old version of the geometry.
The torsion and the basic vector of AP-geometry are also
parameterized and defined by[12],
$$
{{\Lambda}^{\ast}}^{\alpha}_{.\mu \nu}~ \edf~ \nabla^{\alpha}_{.\mu \nu} -\nabla^{\alpha}_{.\nu \mu}~~=~~b~{\Lambda}^{\alpha}_{.\mu \nu}, \eqno{(4.6)}
$$
$$
{C}^{\ast}_{\mu}~~\edf ~~{\Lambda}^{\ast \alpha}_{.\mu
\alpha}~~=~~b~{\Lambda}^{\alpha}_{.\mu \alpha}. \eqno{(4.7)}
$$
The tangent of the new path (4.3) can be written in the form,
$$Z^{\mu}~~=~~U^{\mu}~~+~~b~{\zeta}^{\mu}, \eqno{(4.8)}
$$
where $U^{\mu}$ is the tangent vector of the geodesic of metric and
the vector ${\zeta}^{\mu}$ represents a deviation from geodesic.
The affine parameter $(\tau)$ varying along (4.3) can be related to
that varying along the geodesic $(s)$ by the relation [12],
$$s~=~~\tau ~(1~+~b~ U^{\mu}{\zeta}_{\mu}). \eqno{(4.9)}
$$
For physical reasons [11], the parameter $b$ is suggested to take the form
$$b~=~{\frac{n}{2}}~{\alpha} ~{\gamma},\eqno{(4.10)}
$$
where $n$ is a natural number, $\alpha$ is the fine structure
constant and $\gamma$ is a dimensionless parameter of order
unity.The presence of $\frac{n}{2}$ in the parameter $b$ will
preserve the jumping step appeared in Table II. We are going to call
(4.3) the {\it "Quantum Path Equations"}.

The torsion term, on the R.H.S. of (4.3), is suggested [11] to
represent a type of interaction between the torsion of the
background gravitational field and the quantum spin of the moving
test particle, {\bf Spin-Gravity Interaction}. We are going to
take $n = 0, 1, 2, 3, ....$ for particles with spin $ 0,
\frac{1}{2}, 1 ,\frac{3}{2}, ....$, respectively. For slowly
rotating macroscopic objects, we are going to take $n = 0$.

\section{Quantum Paths and Their Linearization}

The path equation (4.3) can be used as an equation of motion for
any field theory, constructed in the AP-geometry, provided that
the theory has good Newtonian limits. In such theories, (e.g. [7],
[13], [14]), the tetrad vectors  $\h{i}_{\mu}$ are considered as
field variables. So, if we write,
$$
\h{i}_{\mu} = \delta_{_i \mu} + \epsilon h_{_i \mu},          \eqno{(5.1)}
$$
where $\epsilon$ is a small parameter, $\delta_{_i \mu}$ is
Kroneckar delta and $h_{i \mu}$ represents deviations from flat
space, then the weak field condition can be fulfilled by
neglecting quantities of the second and higher orders in
$\epsilon$ in the expanded field quantities. For a static field
assumption, we are going to assume the  vanishing of time
derivatives of the field variables.
 The vector components $Z^\mu$ ($ {\edf \tder{x^{\mu}}{\tau}}$)will have the
 values,
$$
Z^1 \approx Z^2 \approx Z^3 \approx \varepsilon {\  }{\  }{\  }
, Z^0 \approx 1 - \varepsilon ,                 \eqno{(5.2)}
$$
where $\varepsilon$ is a  parameter of the order $( \frac{v}{c})$.
If we want to add the condition of slowly moving particle to the
previous conditions we should neglect quantities of second and
higher orders of the parameter $\varepsilon$. Thus, in expanding
the quantities of the path equation (4.3) we are going to neglect
quantities of orders $\epsilon^2, {\ }\varepsilon^2,{\  } \epsilon
\varepsilon$ and higher, and also time derivatives of the field
variable are to be neglected. To the first order of the
parameters, the only field quantities that will contribute to the
path equation (4.3) are given by [11],
$$
\Lambda_{00}^{.\ .\ i} = - \epsilon h_{ 0 0,  i}   ,{\  }{\  }{\  }{\  }  (i= 1,2,3)
                                                     \eqno{(5.3)}
$$
$$
\{^{\ i}_{0{\ }0}\} = \frac{\epsilon}{2} Y_{00,i}   ,{\  }{\  }{\  }{\  }  (i= 1,2,3)
                                                     \eqno{(5.4)}
$$
where $Y_{\mu \nu}$ is defined by,
$$
g_{\mu \nu}  = \eta_{\mu \nu} + \epsilon {\ } Y_{\mu \nu} {\ }{\ }{\ },
$$
$g_{\mu \nu}$ is given by (2.5) and $ \eta_{\mu \nu}$ is the Minkowski metric tensor . Substituting from (5.3),(5.4) into (4.3)
 we get, after some manipulations :
$$
\frac{d^2x^i}{d{\tau}^2} = -\frac{1}{2}{\ }\epsilon{\ }(1 -\frac{n}{2}\alpha \gamma)
Y_{00,i}{\ }Z^0{\ }Z^0.
                                                         \eqno{(5.5)}
$$
In the present case, the metric of the Riemannian space,
associated to AP-space, can be written in the form [11],
$$
(\frac{d\tau}{dt})^2 = c^2{\ }(1 + \epsilon {\ }Y_{00}).     \eqno{(5.6)}
$$
Substituting from (5.6) into (5.5) we get after some manipulations:
$$
\frac{d^2x^i}{dt^2} = - \frac{c^2}{2}{\ }\epsilon{\ }(1 -\frac{n}{2}\alpha \gamma)
{\ }Y_{00,i}{\ }{\ }{\ }{\ }(i=1,2,3)
$$
which can be written in the form,
$$
\frac{d^2x^i}{dt^2} = - \pder{\Phi_s}{x^i} {\ }{\ }{\ }{\ }(i=1,2,3) {\ }{\ },
                                                   \eqno{(5.7)}
$$
where
$$
\Phi_s \edf \frac{c^2}{2}{\ }\epsilon{\ }(1 -\frac{n}{2}\alpha \gamma)
{\ }Y_{00}.                                                  \eqno{(5.8)}
$$
Equation (5.7) has the same form as Newton's equation of motion of
a particle in a gravitational  field having the potential $\Phi_s$
given by (5.8), which differs from the classical Newtonian
potential.
 In the case of motion of macroscopic  particles $(n=0)$, we get from (5.8):
$$
\Phi_s =  \frac{c^2}{2}{\ }\epsilon{\ }Y_{00} = \Phi_N               \eqno{(5.9)}
$$
where $\Phi_N$ is the Newtonian gravitational potential obtained from a similar
treatment of the geodesic equation. Thus (5.8) can be written in the form,
$$
\Phi_s =  (1 - {\frac{n}{2}} \alpha \gamma) \Phi_N.                  \eqno{(5.10)}
$$
This last expression shows that the gravitational potential felt
by the spinning particle is less than that felt by a spinless
particle or a macroscopic test particle. In other words, the
Newtonian potential is reduced, for spinning particles, by a factor
$(1 - {\frac{n}{2}} \alpha \gamma)$.

\section{Confirmation and Applications of the Quantum Paths}

In the context of geometerization of physics, path equations of an
appropriate geometry, are used to represent trajectories of test
particles. It appears clearly, from the previous section, that in
the case of a static weak field and a slowly moving test particle,
we get Newtonian motion, provided that the particle is spinless. In
the following subsections, we are going to use the quantum path
equation (4.3), and its linearization consequences, to study the
motion of spinning test particles in gravitational fields.

\subsection{The COW-Experiment}
Colella, Overhauser and Werner suggested and carried out
experiments concerning the quantum interference of thermal
neutrons [15], [16], [17]. This type of experiments is known, in
the literature, as the COW-experiment. The aim of the experiment
is to test the effect of the Earth's gravitational field on the
phase difference between two beams of thermal neutrons, one is
more closer to the Earth's surface than the other. The second
version of the COW experiment was carried out by Werner et
al.[18]. This version is characterized by a high accuracy in the
measurements of the phase shift (1 part in 1000). The measurements
show that the experimental results are lower than the theoretical
calculations (using the Newtonian gravity) by about 8 parts in
1000. This is a real discrepancy, which may indicate the presence
of a type of non-Newtonian effects.

Now  one can use equation (4.3) to give an interpretation for the discrepancy in
the COW-experiment. In fact we are going to use the consequence of equation (4.3)
given by equation (5.10)
since the following conditions, under which (5.10) is derived, hold:\\
-Thermal neutrons can be considered as $\underline{slowly}$ moving test particles, and \\
-the Earth's gravitational field can be considered as $\underline{weak}$ and
$\underline{static}$.

The phase difference $(\Delta \Omega)$ between the two beams of
neutrons in the COW-experiment is given by (cf. [19]),
\begin{equation}
{(\Delta \Omega)_N = - {\frac{1}{\hbar}} \int^{ABD}_{ACD} \Phi_N dt ,}
\end{equation}
where ABD and ACD are the trajectories of the upper and lower
beams of neutrons, in the interferometer, respectively . The index
$N$ is used to indicate that (6.1) is obtained using the Newtonian
potential $\Phi_N$, and $\hbar$ is the Planck's constant. Since
neutrons are spinning particles they will be affected by the
torsion of space-time, as suggested. Thus we replace $\Phi_N$ in
(6.1) by $\Phi_S$ given by (5.10). In this case (6.1) will take
the form [20]:
\begin{equation}
{(\Delta \Omega)_S = - (1 - {\frac{n}{2}} \gamma \alpha) {\frac{1}{\hbar}}
\int^{ABD}_{ACD} \Phi_N dt ,}
\end{equation}
i.e.,
\begin{equation}
{(\Delta \Omega)_S = (\Delta \Omega)_N - {\frac{n}{2}} \gamma \alpha
(\Delta \Omega)_N .}
\end{equation}
The index $S$ is used to indicate that (6.2) is obtained using the
potential $\Phi_S$. Taking the value of $\alpha =
{\frac{1}{137}}$, $n = 1$ for spin ${\frac{1}{2}}$-
particles (neutrons), we easily get the following results [20]:\\
(1) the theoretical value of the COW-experiment will decrease by about
4 parts in 1000, if we take $\gamma = 1$, \\
(2) the theoretical value will coincide with the experimental one if we take $\gamma = 2$.

\subsection{SN1987A}
Carriers of astrophysical information are massless spinning
 particles. These carriers are photons, neutrinos, and expectedly,
 gravitons. These three types of carriers are assumed to be
 emitted from supernovae events. In February, $23^{rd}$,1987 a
 supernova , in the Large Magellanic Cloud, was observed (cf.[21]).
 Observations of the arrival time of neutrinos, at the
 Kamiokande detectors, was recorded  in February $23^{rd}$, 1987,
 $7^{h} 35^{m} UT$, while the arrival time of photons was
 on the same day at $10^{h} 40^{m} UT$. The bar of the
 gravitational waves antennae in Rome and Maryland recorded relatively large
 pulses, 1.2 seconds earlier than neutrinos (cf.[22], [23]).
 Although the three types of particles have different spins, general relativity
 assumes that they follow the same trajectory (null-geodesic of the metric), since
 they are all massless.

In the context of general relativity, it is well known that the
time interval required for a massless particle to traverse a given
distance is longer in  the presence of gravitational field having
the potential $\Phi(r)$. The time delay is given by (cf.[24])
\begin{equation}
{\Delta t_{GR} = const.~~ \int_e^a \Phi(r) dt}
\end{equation}
where $e$ and $a$ are the emission and arrival times of  the
 carrier,  respectively.
In SN1987A's time delay (cf. [24], [25], [26]), $\Phi(r)$ is taken
to be the Newtonian potential $\Phi_{N}$ (spin independent). In
this case we can construct a spin-independent model, for the
emission times of the carriers. If we assume that $\Phi(r)$ is the
spin dependent gravitational potential $\Phi_{s}$ (5.10), we then
get the spin-dependent model. The results of these two models [27]
are summarized in table IV.
\\
 \begin{center}
{Table IV: Emission Times Given By The Models}
%\vspace{0.5cm}
\begin{tabular}{|c|c|c|} \hline
 Particles Emitted  & Spin Independent Model  & Spin
Dependent Model
\\ (Cause)&(Null-Geodesic) &(Quantum Path) \\
%& & \\
 \hline & & \\ Neutrino (core collapse)     & 0.0 & 0.0 \\
 % & & \\
\hline & & \\ Photons (maximum brightness) &  $ +3^h 5^m$   &
$+15^h 18^m$ \\
% & & \\
\hline
 & & \\ Gravitons (?) &$-1^s.2$
&$+36^h 28^m$ \\
% & & \\
 \hline
\end{tabular}
\end{center}
\section*{}

From Table IV, we can conclude that, the two models assume two
different scenarios for the emission of carriers of astrophysical
information. {\it {The spin-independent model}} shows an
indication that neutrinos were emitted due to core collapse,
associated with gravitons as a result of sudden change in the
space-time symmetry, probably, due to a kicked born neutron star.
About three hours later photons were emitted as a result of
maximum brightness of the envelope. {\it {The spin-dependent
model}} shows that: neutrinos were emitted due to a core collapse,
preserving sphericity of the core. After 15 hours photons were
emitted due to maximum brightness of the envelope, in agreement
with SN theories, then 21 hours later the envelope explodes
asymmetrically producing a sudden change of space-time symmetry
which causes the emission of gravitational waves. It could be seen
that {\it {the spin-dependent model}} is more preferable than {\it
{the spin-independent model}}.

\subsection{The Cosmological Parameters}
Cosmological information are usually carried by, and extracted
from, massless spinning particles, {\it "carriers of cosmological
information"}. The photon (spin 1-particle) is a good candidate
representing one type of these carriers. Recently, the neutrino
(spin $\frac{1}{2}$-particle) entered the playground as another
type. We expect, in the near future, that a third type of
carriers, the graviton (spin 2-particle), to be used for
extracting cosmological information. Two factors affect the
properties of these carriers. The first is the source of the
carrier. The second factor is the trajectory of the carrier, in
the cosmic space, from its source to the receiver. The first
factor implies the information carried, which reflect the
properties of the source. The second factor represents the impact
of the cosmic space-time on the properties of the carrier. So,
information carried by these particles contain, a part connected
to their sources, and another part related to the space-time
through which these particles travelled.

Cosmological parameters are quantities extracted from the
information carried by the above mentioned particles.
Consequently, the values of such parameters are certainly affected
by the second factor. In the present work, we are going to explore
the impact of this factor on these parameters.

It is well known that, the red-shift of spectral lines, coming
from distant objects, plays an important role in measuring the
cosmological parameters. Theoretical calculations of the
red-shift, in the context of GR, treats it as a metric phenomena,
since the metric of space-time is the first integral of the
geodesic equation. But, if the trajectories of test particles, the
carriers, are spin-dependent, then the red-shift of spectral lines
is no longer a metric phenomena. In this case one should look for
an alternative scheme for calculating this quantity.

Kermack, McCrea and Whittaker
 [28] developed two theorems on null-geodesics which were applied
 to get the standard red-shift of  relativistic cosmology, using the following formula,
 $$ \frac{\lambda_{o}}{\lambda_{1}}~~= ~~\frac {^{1}\eta^{\mu}\rho_{\mu}}
 {^{0}\eta^{\mu}\varpi_{\mu}},   \eqno{(6.5)}$$
where $^{1}\eta^{\mu}$ is the transport vector along the
null-geodesic $\Gamma$ connecting two observers A and B, evaluated
at A, $^{0}\eta^{\mu}$ is the transport vector evaluated at B,
$\rho^{\mu}$ is the unit tangent along the trajectory of A,
$\varpi^{\mu}$ is the unit tangent along the trajectory of B,
$\lambda_{1}$ is the wave length of the spectral line as measured
at A, $\lambda_{o}$ is the wave length of the spectral line as
measured at B, and $\Gamma$ represents the trajectory of a
massless particle from A (source) to B (receiver). If the universe
is expanding then $\lambda_{o} > \lambda_{1}$. It can be shown
that the two theorems, mentioned above, are applicable to any
null-path. So, they can be used for massless spinning particles
following the trajectory (4.3).

In order to evaluate the red-shift using (6.5) one has to know first
the values of the vectors used in this formula. Such vectors are
obtained as solution of the spin-dependent path equation (4.3).
Robertson [29] constructed two geometric AP-structures for
cosmological applications. Using one of these structures, and
performing the necessary calculations we get [30], $$
\frac{\lambda_{o}}{\lambda_{1}}~=~(\frac{R_{o}}{R_{1}})^{(1-\frac{n}{2}
\al \gamma)} \eqno{(6.6)} .$$ Now, we define the spin-dependent
scale factor as,

$${R}^{\ast}
 = R^{(1-\frac{n}{2} \al \g))}. \eqno{(6.7)}$$
Using ${R}^{\ast}$, in place of\ R \ in the standard definitions of
the cosmological parameters, we can list the resulting
spin-dependent parameters in Table V. The second column of this
Table, gives the values of the parameters as if they are extracted
from massless spinless particles. The values of the parameters
extracted from photons should match the values listed in column 4.
It is worth of mention that the matter parameter is not affected by
the spin-gravity interaction. This is due to its independence of
Hubble's parameter.

\begin{center}
 Table V: Spin-Dependent Cosmological Parameters     \\
%\vspace{0.5cm}
\begin{tabular}{|c|c|c|c|c|} \hline
& & & & \\ Parameter & Spin-0  & Spin-$\frac{1}{2}$ (neutrino)&
Spin-1 (photon) & Spin-2 (graviton)
\\ & & & &
\\ \hline & & & & \\ Hubble     & $H_{o}$ & $(1- \frac{\al}{2}) H_o $ & $(1-\al)H_{o}$
&  $(1- 2 \al)H_{o}$ \\ & & & & \\ \hline & & & & \\
 Age &  $\tau_{o}$ & $\frac{\tau_o}{(1-\frac{\al}{2})}$ &
$\frac{\tau_{o}}{(1-\al)}$ &  $\frac{\tau_{o}}{(1- 2 \al)}$ \\
 & & & & \\ \hline & & & & \\ Acceleration & $A_o$
 & $(1-\frac{\al}{2})(A_{o}-\frac{\al}{2}H_{o})$
 & $(1-{\al})(A_{o}-{\al}H_{o})$
 & $(1-2{\al})(A_{o}-2{\al}H_{o})$
\\ & & & & \\ \hline & & & & \\ Deceleration  &$ q_{o}$
&$\frac{(q_{o}-\frac{\al}{2H_{o}})}{(1-\frac{\al}{2})}$ &
$\frac{(q_{o}-\frac{\al}{H_{o}})}{(1-\al)}$ &
$\frac{(q_{o}-\frac{2 \al}{H_{o}})}{(1- 2\al)}$
\\ & & & & \\ \hline
\end{tabular}
\end{center}

There are some evidences for the existence of the spin-gravity
interaction on the laboratory scale (the results of the
COW-experiment), and on the galactic scale  (the data of SN1987A).
Now, to verify the existence of this interaction on the
cosmological scale, observations of one parameter at least, using
two different types of carriers, are needed. For example, if we
observe neutrinos and photons to get Hubble's parameter, a
discrepancy of order 0.001 would be expected, if this interaction
exists on the cosmological scale.

\section{General Discussion and Concluding Remarks}
In the present work, it is shown that, starting within the
geometerization philosophy, some quantum properties appeared very
naturally in the structure of two types of non-symmetric
geometries (see the third column of Tables II and III). These
properties emerged without imposing any known quantization schemes
on the geometry. The properties characterize the torsion term of
two new sets of path equations discovered in each geometry, (2.25)
and (2.45). The natural appearance of such properties can be
considered as quantum roots built in non-symmetric geometry.

It is shown that these roots could be extended and diffuse in the
whole geometry, using a certain parameterization scheme, suggested
in Section 4. This scheme, applied to the AP-geometry, could be
applied with some efforts to the ENS-geometry. The application of
the parameterization scheme, not only diffuses the quantum
properties in the whole geometric structure, but also solves the
two main problems of the AP-geometry, mentioned in Subsection 2.1.
We can summarize the main advantages of this scheme in the
following points:

1. As stated above, it extends the quantum roots, appeared in the
path equations, of a non-symmetric geometry, to other geometric
entities.

2. It solves completely the curvature problems (the identical
vanishing of the curvature (2.12)) , mentioned in Subsection 2.1,
by defining a general parameterized non-vanishing curvature tensor
(4.4).

3. From the application point of view and depending on the
curvature (4.4), field theories written in the parameterized
AP-geometry do not need the condition for a vanishing torsion
(which leads to a flat AP-space via (2.18-21)), in order to get a
correct GR-limit. In other words, to switch-off the effect of the
torsion, in such theories, we only take the parameter $b = 0$.

4. It solves the second problem of conventional AP-geometry, i.e.
the non-physical applicability of the path equations (2.13). The
new quantum paths (4.3) could be used for physical applications,
as shown in Section 6.

5. The parameterized absolute parallelism (PAP) geometry is more
general than both the Riemannian and conventional AP-geometries.
It could account for both geometries as two limiting cases. These
limits can be obtained using (4.1b). The first limit $a = 0
\Longrightarrow b = 1$, which corresponds to the conventional
AP-geometry. The second is $a = 1 \Longrightarrow b = 0$, which
corresponds to the Riemannian geometry. Figure 1 [12], is a
schematic diagram giving the complete spectrum of geometries
admitted by the PAP-geometry.
\begin{center}
\includegraphics[width=10cm]{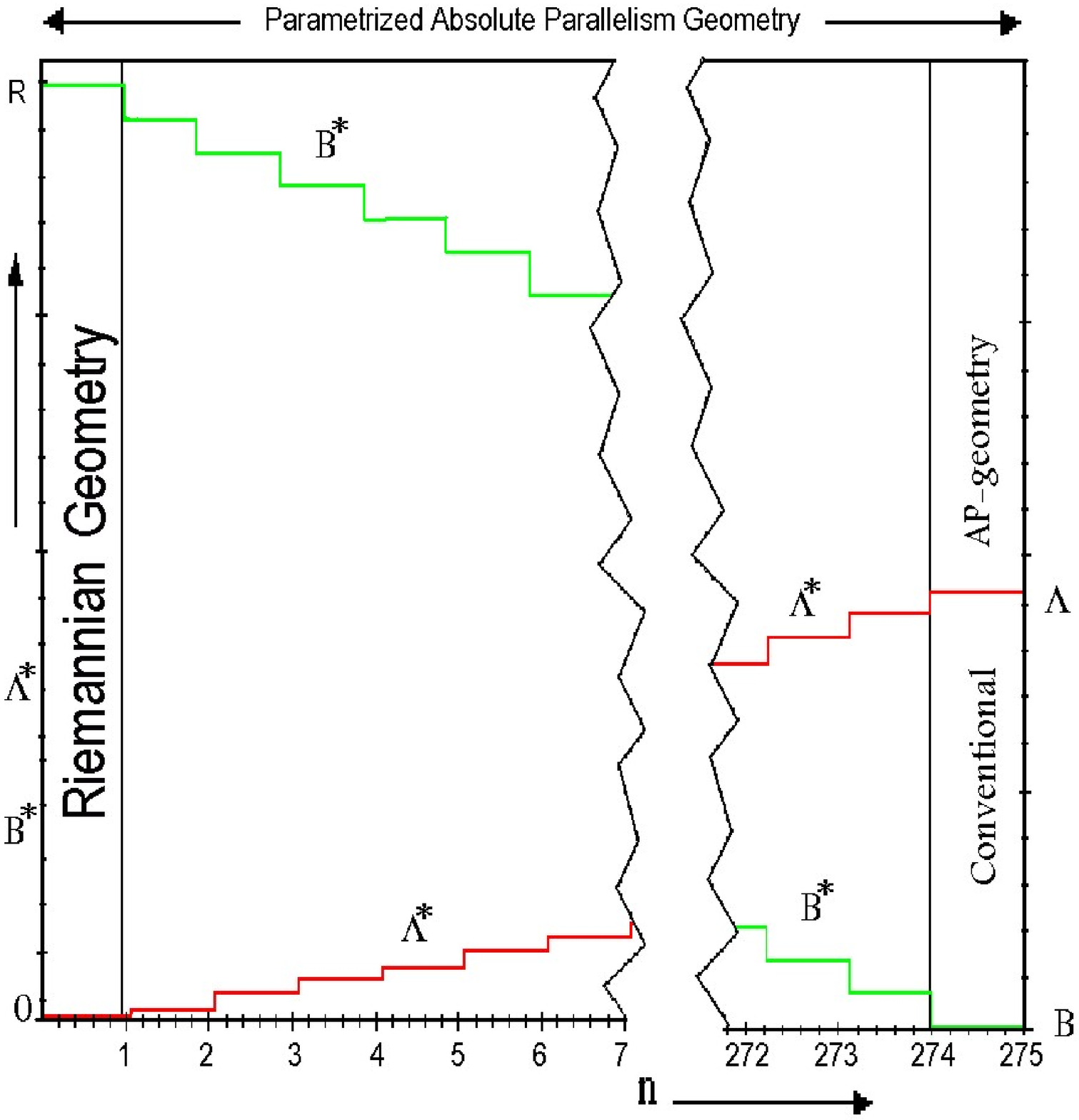}\\[5pt]
%\caption{Quantum Properties of PAP-Geometry}
\end{center}
\begin{quote}
\begin{center}
\textbf{Figure 1:} Quantum Properties of PAP-Geometry.
\end{center}
\end{quote}

This Figure is plotted using equations (4.4) and (4.6).\\

The new quantum paths (4.3)can be reduced to the geodesic equation
of Riemannian geometry (or null-geodesic upon reparameterization),
upon setting $b = 0$, which switch-off the effect of the torsion
term. In this case the equation can account for classical
mechanics and relativistic mechanics. But if $b \neq 0$, then the
torsion of the background gravitational field will interact with
some of the properties of the moving particle. Recalling that the
parameter $b$ jumps by steps of one-half, (4.10), then one can
conclude that the property of the test particle, by which it
interacts with the torsion, is its quantum spin. For this reason,
the torsion term in (4.3) is suggested to represent "Spin-Gravity
Interaction". The linearization carried out in Section 5, shows
clearly that this interaction will reduce Newton's gravitational
potential, as obvious from (5.10). This equation shows that, the
gravitational potential felt by a spinning particle is less than
that felt by a spinless particle, or by a macroscopic object. In
other words, one can say that, spinning particles feel the
space-time torsion. This is similar to the fact that charged
particles feel the electromagnetic potential, while neutral
particles do not feel it.\\

The discrepancy, between the experimental results and the
theoretical calculations (using Newtonian gravity), of the
COW-experiment gives a good indicator for the existence of
spin-gravity interaction, on the laboratory scale. The experimental
results are found to be lower than the expected theoretical
calculations. This discrepancy can be interpreted, qualitatively, by
a decrease in the gravitational potential, of the Earth, felt by
neutrons (spin one-half particles). The value of this potential,
felt by neutrons, is less than the value given by Newton's theory.
The application of the new quantum path (4.3), Subsection 6.1, gives
good, qualitative and quantitative, agreement with the experimental
results. Such agreement gives, not only an evidence of the existence
of spin-gravity interaction, but
also a direct confirmation of equation (4.3).\\

The application of the linearized form of (4.3), in the case of
motion of spinning massless particles coming from SN1987A,
Subsection 6.2, gives a good model for the emission times of these
particles from this supernova (see Table IV). This may indicate
the presence of the spin-gravity interaction on the galactic
scale. But more efforts are still needed, both to confirm
supernovae mechanisms and for observing more supernovae, to give
strong confirmation for the existence of this interaction on the
astrophysical scale.\\

The full path equation (4.3) is applied in the case of cosmology,
Subsection 6.3. It is shown that the values of the cosmological
parameters will be affected by the spin-gravity interaction, if it
exists on the cosmological scale. The values of these parameters
will depend on the spin of the particle, from which cosmological
information are extracted. It is suggested that, a cosmological
parameter measured using two massless particles, with different
spins (e.g. photon and neutrino) may confirm the existence of
spin-gravity interaction on the cosmological scale. The
sensitivity of the apparatus, or experiment, to be used should be
better than $0.001$\\

In view of the present work, I will try to give short probable
answers to some of the good questions raised by professor V.Petrov
in the closing session of the conference:

Q1: What is the appropriate topology/geometry?

A1: A non-symmetric geometry.

Q2: How many dimensions?

A2: So for, in the context of geometerization of physics, we don't
need more that four dimensions. Mass and charge appear as
constants of integration. There are some attempts to represent
other interactions (e.g. electromagnetism) together with gravity
in spaces of four dimensions (cf. [7]).

 Q3: What are the
experimental/observational signature of quantum-geometrical
effects?

A3: Concerning the experimental signature, the COW-type experiment
is a good media for testing quantum-geometrical effects on the
laboratory scale. the discrepancy in the results of this
experiment gives a good indicator for the existence of such
effects.

Concerning the observational signature, more efforts are needed
for observing photons and neutrinos (and probably gravitons, in
the future), from supernovae events, in order to detect the
existence of such effects on the astrophysical and cosmological
scales.\\

Finally, I would like to thank the organizing committee and
Professor V.Petrov for inviting me to participate in the conference
and to give this talk.\\ \\ \\ \\
{\bf References}\\
{ [1] Eisenhart, L.P. (1926) {\it "Riemannian Geometry"}, Princeton Univ. Press}. \\
{ [2] Einstein, A. (1929) Sitz. Preuss. Akad. Wiss., {\bf 1}, 1.} \\
{ [3] Einstein, A. (1955) {\it "The Meaning of Relativity"}, Appendix II, Princeton.}\\
{ [4] Bazanski, S. L., (1977) Ann. Inst. H. Poincar\'e, A{\bf 27}, 145.} \\
{ [5] Bazanski, S. L., (1989) J. Math. Phys., {\bf 30}, 1018}.\\
{ [6] Wanas, M.I.(1975) Ph.D. Thesis, Cairo University.} \\
{ [7] Mikhail, F.I. and Wanas, M.I. (1977) Proc. Roy. Soc. London, {\bf A356}, 471.} \\
{ [8] Wanas, M.I., Melek, M. and Kahil, M.E. (1995) Astrophys. Space
Sci., {\bf 228}, 273.;

      gr-qc/0207113} .\\
{ [9] Wanas, M.I. and Kahil, M.E. (1999) Gen. Rel. Grav., {\bf 31}, 1921.; gr-qc/9912007}. \\
{[10] Wanas, M.I. (2000) Turk. J. Phys., {\bf 24}, 473.; gr-qc/0010099}.\\
{[11] Wanas, M.I. (1998) Astrophys. Space Sci., {\bf 258}, 237.; gr-qc/9904019}. \\
{[12] Wanas, M.I. (2002) Proc. MG IX, Vol. 2, 1303.} \\
{[13] M\o ller, C. (1978) Math. Fys. Medd. Dan.Vid. Selsk., {\bf 39}, 1.} \\
{[14] Hayashi, K. and Shirafuji, T. (1979) Phys. Rev. D{\bf 19}, 3524.} \\
{[15] Overhauser, A.W. and Colella, R. (1974) Phys. Rev. Lett., {\bf 33}, 1237.}\\
{[16] Colella, R., Overhauser, A.W. and Werner, S.A.(1975) Phys. Rev. Lett., {\bf 34},1472.}\\
{[17] Staudenmann, J.L., Werner, S.A., R. Colella, R. and A. W.
Overhauser, A.W. (1980)

      Phys.Rev. A, {\bf 21}, 1419.}\\
{[18] Werner, S.A.,  Kaiser, H.,  Arif, M. and Clother, R. (1988) Physica B, {\bf 151}, 22.}\\
{[19] Greenberger, D.M. (1983) Rev. Mod. Phys., {\bf 55}, 875.}\\ \\
{[20] Wanas, M.I., Melek, M. and Kahil, M.E. (2000) Gravit. Cosmol.
{\bf 6}, 319;

      gr-qc/9812085}. \\
{[21] Schramm, D.N. and Truran, J.W. (1990) Phys. Rep. {\bf 189}, 89-126.} \\
{[22] Weber, J. (1994) Proc. First Edoarndo Amaldi Conf.{\it "On
gravitational wave

      experiment"} Ed. E.Coccia et al. World Scientific P.416.}\\
{[23] De Rujula, A. (1987) Phys. Lett. {\bf 60}, 176.} \\
{[24] Krauss, L.M. and Tremaine, S. (1988) Phys. Rev. Lett., {\bf 60}, 176.} \\
{[25] Longo, M.J. (1987) Phys. Rev. D, {\bf 36}, 3276.} \\
{[26] Longo, M. J. (1988) Phys. Rev. Lett., {\bf 60}, 173.}\\
{[27] Wanas, M.I., Melek, M. and Kahil, M.E. (2002) Proc. MG IX, Vol
2, 1100;

      gr-qc/0306086}. \\
{[28] Kermack, W.O., McCrea, W.H. and Whittaker, E.T. (1933) Proc.
Roy.

Soc. Edin., {\bf 53}, 31.}\\
{[29] Robertson, H.P. (1932) Ann. Math. Princeton (2), {\bf 33}, 496.} \\
{[30] Wanas, M.I. (2002) To appear in the Proc. IAU-Symp.\# 201,
held in Manchester,

      August 2000.} \\

\end{document}